\documentclass{ragtime}
\usepackage{graphicx}  % standard LaTeX graphics tool;
\usepackage{amsmath}   % For getting proper math eqns
\usepackage{amssymb}   % Used for getting various symbols eg. gtrsim, lesssim etc.
\usepackage{bm} % For bold math (esp of lower greek letters)
\usepackage{dcolumn}% Align table columns on decimal point
\usepackage{color}
\usepackage{mathrsfs}
\usepackage{ulem}
\hypersetup{
        colorlinks=true,
        allcolors=blue,
        }
\usepackage{txfonts}
\usepackage{amsfonts}
\usepackage[utf8]{inputenc}
\usepackage{varioref}
\usepackage{csquotes}
\usepackage{dirtytalk}
\usepackage[caption=false]{subfig}

\labelformat{section}{Section-(#1)} 
\labelformat{subsection}{Section-(#1)} 
\labelformat{subsubsection}{Section-(#1)}
\labelformat{subsubsubsection}{Section-(#1)}
\labelformat{figure}{Fig.~(#1)} 
\labelformat{equation}{Eq.~(#1)}
\labelformat{subfigure}{Fig.~(\thefigure#1)} 
\labelformat{table}{Tab.~#1} 
\labelformat{appendix}{Appendix #1}

\title{Can extended bodies follow geodesic trajectories?}

\author[G. Lukes-Gerakopoulos and S. Mukherjee]{Georgios Lukes-Gerakopoulos \at[]{1,a} and Sajal Mukherjee\at[]{1,2,b} \\
\ins{1}  Astronomical Institute of the Czech Academy of Sciences, Bo\v{c}n\'{i} II 1401/1a,\splitins[1] CZ-14100 Prague, Czech Republic \\
\ins{2}  Department of Physics, Birla Institute of Technology and Science - Pilani, Rajasthan 333031, India \\
\ins{a}\Email{gglukes@gmail.com}\\
\ins{b}\Email{sajal.mukherjee@pilani.bits-pilani.ac.in}}

\begin{document}

\begin{abstract}
We provide an extension of the analysis on whether an extended test body can follow a geodesic trajectory given by \bibentry{Mukherjee2022}. In particular, we consider a test body in a pole-dipole-quadrupole approximation under the Ohashi-Kyrian-Semer\'{a}k spin supplementary condition moving in the Schwarzschild and Kerr background. Using orbital setups under which a pole-dipole body can follow geodesic motion, we explore under which conditions this can take place also in the pole-dipole-quadrupole approximation, when only the mass quadrupole is taken into account. For our analysis we employ the assumption that the dipole contribution and the quadrupole contribution vanish independently.   
\end{abstract}

\section{Introduction} \label{sec:intro}

The framework of an extended test body moving on a curved background, which we employ, dates back to the pioneering works of \citet{Mathisson:1937zz} and \citet{Papapetrou:1951pa}. A significant contribution to this framework was provided by \citet{dixon1970dynamics,dixon1970dynamics2,dixon1974dynamics}, which resulted in calling the respective equation of motion the Mathisson-Papapetrou-Dixon (MPD) equations. These equations in the pole-dipole-quadrupole approximation, when only gravitational interactions are considered, read \citep{Steinhoff:2009tk,Steinhoff:2012rw}
\begin{align}
\dot{P}^{\mu} &= -\frac{1}{2}R^{\mu}_{~\nu \alpha \beta}\mathcal{U}^{\nu} S^{\alpha \beta}-\dfrac{1}{6}J^{\alpha \beta \gamma \delta} \nabla^{\mu}R_{\alpha \beta \gamma \delta}, \label{eq:MPD_P}  \\
\dot{S}^{\mu \nu} &= 2 P^{[\mu} \mathcal{U}^{\nu]}+\dfrac{4}{3} J^{\alpha \beta \gamma [\mu}R^{\nu]}_{~\gamma \alpha \beta },\label{eq:MPD_S}
\end{align}
where $R^{\mu}_{~\nu \alpha \beta}$ is the Riemann tensor, $S^{\alpha \beta}$ is the spin tensor, $J^{\alpha \beta \gamma \delta}$ is the quadrupole tensor, $P^{\mu}$ is the four-momentum  and $\mathcal{U}^{\mu}$ is the four-velocity, while the \enquote*{dot} defines a covariant derivative with respect to the proper time. The MPD equations do not describe the evolution of the quadrupole moment, which has to be determined by the matter structure of the body \citep[see, e.g.,][]{Steinhoff:2012rw}.

It is obvious from \ref{eq:MPD_P} that an extended body should in general deviate from a geodesic trajectory even if $P^\mu||\mathcal{U}^\mu$. Note that $P^\mu||\mathcal{U}^\mu$ is not in general the case for the MPD equations. The question we tackled in \citep{Mukherjee2022} was whether these rules can have exceptions on a black hole background, i.e. can extended bodies follow geodesic trajectories? We have found some positive answers in the pole-dipole approximation, but failed to do so in the pole-dipole-quadrupole in the spin induced quadrupole case.

In the present work, we restrict our analysis to the contribution of the mass quadrupole moment $Q^{\beta\gamma}$ to the quadrupole term. Then, the quadrupole tensor reads
\begin{equation}\label{eq:QuadMassOKS}
J^{\alpha \beta \gamma \delta}=-3 \mathcal{V}^{[\alpha}Q^{\beta][\gamma} \mathcal{V}^{\delta]},
\end{equation}
where $ \mathcal{V}$ is a future oriented time-like vector, which is employed to fix the centre of the mass of the body by using the constraint 
\begin{align} \label{eq:SSC}
 \mathcal{V}_\mu S^{\mu\nu}=0.
\end{align}
This constraint is known in the literature as the spin supplementary condition (SSC). There are several SSCs \citep[see, e.g.,][]{Costa:2017kdr}, but in \citep{Mukherjee2022} as in this work we focus on the Ohashi-Kyrian-Semer\'{a}k (OKS) one \citep{Ohashi03,Kyrian:2007zz}. For this SSC a vector $\mathcal{V}^\mu=w^\mu$ is chosen so that $\dot{w}^\mu=0$ and $w^\mu w_\mu=-1$. 

This OKS choice leads to $\dot{S}^{\mu\nu}w_\mu=0,~\ddot{S}^{\mu\nu}w_\mu=0$ and all the contractions of $w^\mu$ with the higher covariant derivatives of the spin tensor are also equal to zero. Hence, by contracting \ref{eq:MPD_S} with $w_\mu$ leads to
\begin{equation}\label{eq:MomentumOKS}
    P^\mu=\frac{1}{-w_\nu \mathcal{U}^\nu}[(-P^\gamma w_\gamma) \mathcal{U}^\mu+K^{\mu\delta} w_\delta],
\end{equation}
where we have set $K^{\mu \nu}=\dfrac{4}{3} J^{\alpha \beta \gamma [\mu}R^{\nu]}_{~\gamma \alpha \beta }$. It is straightforward that if there is no quadrupole term, then the four-momentum and the four-velocity become parallel, which a feature of geodesic orbits. \citet{Mukherjee2022} have shown that this feature of OKS SSC can be recovered even with the quadrupole term, as long as
\begin{align}\label{eq:PDQMomVan}
    K^{\mu\gamma} w_\gamma=0
\end{align}
holds.

In the cases that $P^\mu||\mathcal{U}^\mu$, it is said that the hidden momentum vanishes, since $P^\mu$ can be split in a part parallel $P_{\|}^\mu$ to $\mathcal{U}^\mu$ and to the hidden momentum part $P^{\mu}_{\rm hid}$ \citep{Costa:2014nta} leading to
\begin{equation}\label{eq:MomDec}
    P^\mu=P_{\|}^\mu+P^{\mu}_{\rm hid}
\end{equation}
The parallel part reads
\begin{equation}\label{eq:MomPar}
    P_{\|}^\mu=m\, \mathcal{U}^\mu ,
\end{equation}
where the mass $m:=-P_\mu \mathcal{U}^\mu$. In the MPD formalism there is also another mass defined as $\mu^2=-P^\mu P_\mu$. Note that $\mu$ and $m$ in general do not coincide.

Our work concerns motion in a Kerr black hole background, for which the metric tensor in the Boyer-Lindquist coordinates $\{t,r,\theta,\phi\}$ reads
 \begin{align} \label{eq:KerrMetric}
   g_{tt} &=-\left(1-\frac{2 M r}{\Sigma}\right) \; ,\:
   g_{t\phi} = -\frac{2 a M r \sin^2{\theta}}{\Sigma} \; ,\:
   g_{\phi\phi} = \frac{(\varpi^4-a^2\Delta \sin^2\theta) \sin^2{\theta}}{\Sigma} \; , \nonumber\\
   g_{\theta\theta} &= \Sigma \; , \:
   g_{rr} = \frac{\Sigma}{\Delta} \; ,
 \end{align} 
with
 \begin{align}
  \Sigma = r^2+ a^2 \cos^2{\theta} \; ,\:
  \Delta = \varpi^2-2 M r \; ,\:
  \varpi^2 = r^2+a^2 \; , \label{eq:Kerrfunc} 
 \end{align}
where $M$ is the mass of the black hole and $a$ corresponds to its angular momentum per mass $M$. By setting $a=0$ in Kerr's metric tensor, we recover the spherically symmetric Schwarzschild black hole metric tensor. To simplify some computations in Kerr \citep{Carter:1968ks}  we can employ the following tetrad field
\begin{eqnarray}
e^{(0)}_{\mu} & = & \left(\sqrt{\dfrac{\Delta}{\Sigma}},0,0,-a \sin^2\theta \sqrt{\dfrac{\Delta}{\Sigma}} \right), \quad
e^{(1)}_{\mu}  =  \left(0,\sqrt{\dfrac{\Sigma}{\Delta}},0,0\right) \nonumber, \\
e^{(2)}_{\mu} & = & \left(0,0,\sqrt{\Sigma},0,0 \right), \quad
e^{(3)}_{\mu}  =  \left(\dfrac{-a \sin\theta}{\sqrt{\Sigma}},0,0,\dfrac{r^2+a^2}{\sqrt{\Sigma}}\sin\theta \right).
\label{eq:Tetrad}
\end{eqnarray}

Having briefed the basic concepts and notions for our analysis. We will start our investigation with the simpler Schwarzschild case and then move to the more complex Kerr one. In both of these investigations we make the assumption that the dipole and the quadrupole contribution vanish independently. To address a more realistic scenario, this assumption should be dropped. However, we believe that it allows an interesting insight into the problem. In the framework provided by this assumption, we can use the results of \citet{Mukherjee2022} for the pole-dipole case and then deal with the quadrupole contribution to the MPD equations. For the quadruple contribution, we first tackle the issue whether the hidden momentum vanish and then we explore under which condition a geodesic motion is possible.

\section{Radial motion in the Schwarzschild background} \label{sec:SBH_PDQ}

In the pole-dipole approximation an extended body can move on a geodesic trajectory in the Schwarzschild background, iff it follows a radial trajectory \citep{Mukherjee2022}. The geodesic motion in this spherically symmetric background can be restricted to the equatorial plane, i.e.
\begin{align}
\mathcal{U}_{t}& =-\tilde{E}_{\rm g}, \quad \mathcal{U}_{r}=\pm \sqrt{\tilde{E}_{\rm g}^2-f\left(1+\frac{\tilde{L}^2_{\rm z}}{r^2}\right)} \Big/\left(1-\dfrac{2 M}{r}\right),\quad
\mathcal{U}_\theta= 0, \quad \mathcal{U}_{\phi}=\tilde{L}_{\rm z},
\label{eq:4-velocity_02}
\end{align}
where $\tilde{E}_{\rm g}$ and $\tilde{L}_{\rm z}$ denote the conserved specific geodesic energy and orbital angular momentum,respectively. It is obvious that for the radial motion $\tilde{L}_{\rm z}=0$. Because of the OKS SSC, i.e. $\dot{w}^\mu=0$, the radial motion  implies for the polar component of $w^\mu$ that
\begin{equation}\label{eq:wth_par}
\dfrac{dw^{\theta}}{d\tau}+\dfrac{\mathcal{U}^{r}}{r}w^{\theta}=0,  
\end{equation}
i.e. $r w^{\theta}=\text{constant}$. By assuming for simplicity that this constant is zero, we get $w^{\theta}=0$. On a radial trajectory the azimuthal component of $w^\mu$ obeys a similar equation to \ref{eq:wth_par}, which allows as also to have $w^{\phi}=0$ for simplicity.

Let us now discuss whether the hidden momentum vanishes for the radial trajectory, i.e. if the \ref{eq:PDQMomVan} is satisfied. By applying $w^{\theta}=0$ on \ref{eq:QuadMassOKS} for the OKS SSC we get
\begin{align} \label{eq:torqueS}
K^{tr}&= K^{\theta \phi}=0, \nonumber \\
K^{t\theta} &= \dfrac{9M}{2r^2(2M-r^2)}\bigl\{w^{r}(Q^{\theta r}w^t-Q^{\theta t}w^r)+r(2M-r)w^{\phi}(Q^{\theta \phi}w^{t}-Q^{\theta t}w^{\phi})\bigr\}, \nonumber \\
K^{t\phi}&=\dfrac{9M}{2r^2(2M-r^2)}\bigl\{w^{r}(Q^{ \phi r}w^t-Q^{\phi t}w^r)+w^{\phi}(Q^{rt}w^{r}-Q^{rr}w^{t}) \nonumber  \\
 &+r(r-2M)Q^{\theta \theta}w^{t}w^{\phi}\bigr\}, \\ 
 K^{r \theta}&=\dfrac{9M}{2r^4}\bigl\{r^3 w^{\phi}(Q^{ \theta r}w^{\phi}-Q^{\theta \phi}w^r)+(2M-r)w^{t}(Q^{\theta r}w^t-Q^{\theta t}w^r)\bigr\}, \nonumber \\
K^{r\phi}&=\dfrac{9M(2M-r)}{2r^4}\bigl\{w^t(Q^{\phi r}w^{t}-Q^{\phi t}w^r)+w^{\phi}(Q^{tt}w^r-Q^{tr}w^t)\bigr\}-\dfrac{9Mr^3Q^{\theta \theta}w^{r}w^{\phi}}{2r^4}.\nonumber 
\end{align}
Taking the above relations into account when calculating $K^{\mu\nu}w_\nu$ does not allow the contraction to vanish identically. Only if we assume $w^{\phi}=0$ along with $w^{\theta}=0$ leads to $K^{\mu\nu}w_\nu=0$. Once the latter holds, we have a vanishing hidden momentum, allowing $P^\mu||\mathcal{U}^\mu$. Since we want the body to be on a geodesic trajectory, i.e. $\dot{\mathcal{U}}^\mu=0$, then $\dot{P}^\mu=\dot{m}\mathcal{U}^\mu$. However, as we show below, we can achieve the same result without setting $w^{\phi}=0$ under our assumption of independently vanishing dipole and quadrupole terms. 

The MPD equations in the case of geodesic motion reduce to 
\begin{align} 
    \dot{m}~\mathcal{U}^\mu &=-\frac{1}{2}{R^\mu}_{\nu\kappa\lambda}\mathcal{U}^\nu S^{\kappa\lambda}+F^\mu, \label{eq:MPD_PR} \\
    \dot{S}^{\mu\nu} &=K^{\mu\nu} \label{eq:MPD_SR},
\end{align}
where $F^{\mu}=-\dfrac{1}{6}J^{\alpha \beta \gamma \delta} \nabla^{\mu}R_{\alpha \beta \gamma \delta}$. Since on the radial trajectory for the pole-dipole the spin curvature coupling vanishes and $\dot{S}^{\mu\nu}=0$ \citep{Mukherjee2022}, then we have that: a) $\dot{S}^{\mu\nu}=0$ implies because of \ref{eq:MPD_SR} that $K^{\mu\nu}=0$ and b)
\begin{align}
    \dot{m}~\mathcal{U}^\mu &=F^\mu. \label{eq:MPD_PRR}
\end{align}
We see that assuming independent vanishing of the dipole and quadrupole contributions, we are led to $K^{\mu\nu} w_\nu=0$ without necessarily setting $w^\phi=0$.

For $K^{\mu\nu}=0$, under the reasonable assumptions that in general $w^r \neq 0$ and $w^t \neq 0$, \ref{eq:torqueS} boils down to
\begin{eqnarray}
 Q^{tr}&=& \dfrac{1}{2w^t w^r}(Q^{rr}[w^t]^2+Q^{tt}[w^r]^2)+\dfrac{rQ^{\theta \theta}}{2(r-2M)w^t w^r}\big\{-(r-2M)^2[w^t]^2+r^2[w^r]^2\big\} \nonumber\\
 Q^{t\theta}&=&\dfrac{Q^{r\theta}w^{t}}{w^{r}}, \quad Q^{\theta \phi}=\dfrac{Q^{r\theta}w^{\phi}}{w^{r}},\label{eq:Q_SBH}  \\
 Q^{t\phi}&=& \dfrac{Q^{r\phi}w^{t}}{w^{r}}+\dfrac{w^{\phi}}{2w^t[w^{r}]^2}\Bigl\{Q^{tt}[w^r]^2-Q^{rr}[w^t]^2-\dfrac{rQ^{\theta \theta}[r^2(w^r)^2+(2M-r)^2(w^t)^2]}{2M-r}\Bigr\}. \nonumber
\end{eqnarray}
By substituting the above relations \ref{eq:Q_SBH} into $F^{\mu}$, we obtain 
\begin{align}
    F^r &=\dfrac{3M[r^2(w^r)^2-(r-2M)^2(w^t)^2]}{2r^4(w^r)^2}\Big[\big\{Q^{rr}+(2M-r)Q^{\theta \theta}\big\}(w^{\phi})^2\nonumber\\
    &+(Q^{\phi \phi}-Q^{\theta \theta})(w^r)^2-2Q^{r\phi}w^rw^{\phi})\Big],
\end{align}
and $F^{t}=F^{\theta}=F^{\phi}=0$. The latter implies that $\dot{m}=0$, which in turn implies that $F^r=0$ resulting in the additional constraint
\begin{equation}
Q^{\phi\phi}=\dfrac{1}{[w^{r}]^2}\Big[Q^{\theta\theta}([w^{r}]^2+(r-2M)[w^{\phi}]^2)+w^{\phi}(2Q^{r\phi}w^{r}-Q^{rr}w^{\phi})\Big].    \label{eq:Qphiphi_SBH}
\end{equation}
To sum up, we have imposed geodesic motion in the Schwarzschild spacetime to a pole-dipole-quadrupole body under OKS SSC assuming that only the mass quadrupole contributes to the quadrupole tensor. Moreover, we have assumed that the dipole and the quadrupole contributions to the MPD vanish independently. This setup led us to the constraints~\ref{eq:Q_SBH} and \ref{eq:Qphiphi_SBH}, which are coupling components of the mass quadrupole tensor $Q^{\mu\nu}$ and the components of the reference vector $w^\mu$. The MPD equations do not prescribe an evolution equation for the quadrupole tensor and so from this point of view there is no evolution equation for the mass quadrupole tensor. However, for the reference vector we have $\dot{w^\mu}=0$, hence the constraints~\ref{eq:Q_SBH} and \ref{eq:Qphiphi_SBH} dictate the way the mass quadrupole tensor evolves under OKS SSC, if the pole-dipole-quadrupole body is set to follow a geodesic trajectory.

\subsection{Equatorial motion in the Kerr background} \label{sec:PQK}

In this section we discuss under which conditions a pole-dipole-quadrupole body can follow an equatorial geodesic trajectory in the Kerr background. As in \ref{sec:SBH_PDQ} we  discuss the case for which the mass quadrupole is the only component of the quadrupole moment (\ref{eq:QuadMassOKS}). On the equatorial plane \ref{eq:wth_par} holds also for Kerr \citep{Harms2016}. Hence, we can set $w^\theta=0$ for simplicity, which in turn implies that $w^{(\theta)}=0$. The $ K^{\mu\nu}$ expressed in the tetrad field frame reads 
\begin{eqnarray} \label{eq:torqueK}
 K^{(t)(r)}&=&K^{(\theta)(\phi)}=0,\nonumber \\
 K^{(t)(\theta)}&=&\dfrac{9M}{2r^3}\bigl\{w^{(r)}(Q^{(t)(\theta)}w^{(r)}-Q^{(r)(\theta)}w^{(t)})+w^{(\phi)}(Q^{(\theta)( \phi)}w^{(t)}-Q^{(\theta)(t)}w^{(t)})\bigr\},\nonumber \\
 K^{(t)(\phi)}&=& -\dfrac{9M}{2r^3}\bigl\{w^{(r)}(Q^{(r)(\phi)}w^{(t)}-Q^{(t)(\phi)}w^{(r)})+w^{(\phi)}(Q^{(t)(r)}w^{(r)}-Q^{(r)(r)}w^{(t)})\nonumber \\ 
 &+& Q^{(\theta)(\theta)}w^{(t)}w^{(\phi)}\bigr\},\nonumber \\ 
 K^{(r)(\theta)} &=& \dfrac{9M}{2r^3}\bigl\{w^{(t)}(Q^{(t)(\theta)}w^{(r)}-Q^{(r)(\theta)}w^{(t)})+w^{(\phi)}(Q^{(\theta)(\phi)}w^{(r)}-Q^{(r)(\theta)}w^{(\phi)})\bigr\}, \\
 K^{(r)(\phi)}&=& -\dfrac{9M}{2r^3} \bigl\{w^{(t)}(Q^{(r)(\phi)}w^{(t)}-Q^{(t)(\phi)}w^{(t)})+w^{(\phi)}(Q^{(t)(t)}w^{(r)}-Q^{(t)(r)}w^{(t)})\nonumber \\ 
 &+& Q^{(\theta)(\theta)}w^{(r)}w^{(\phi)}\bigr\}.\nonumber 
\end{eqnarray}
The above equations imply that on the equatorial plane $K^{(\mu\nu)} w_{(\nu)}$ does not vanish if we just set $w^{(\theta)}=0$, we need to set also $w^{(\phi)}=0$ to achieve it. However, since \citet{Mukherjee2022} showed that the spin curvature coupling for a pole-dipole body vanishes for equatorial trajectories satisfying the relation $\tilde{L}_z=a \tilde{E}_g$, by following the same procedure as in \ref{sec:SBH_PDQ}, i.e. by assuming that the dipole and quadrupole contribution vanish independently, we end up with $K^{(\mu)( \nu)}=0$. This results in $K^{(\mu)( \nu)} w_{(\nu)}=0$ and \ref{eq:torqueK} provides the following relations between the components of the quadrupole moment tensor:
\begin{align}
 Q^{(t)(r)}&= \dfrac{Q^{(r)(r)}[w^{(t)}]^2+Q^{(t)(t)}[w^{(r)}]^2+Q^{(\theta)(\theta)}(-[w^{(t)}]^2+[w^{(r)}]^2)}{2w^{(t)}w^{(r)}}, \nonumber  \\
 Q^{(t)(\theta)}&= \dfrac{Q^{(r)(\theta)}w^{(t)}}{w^{(r)}}, \quad Q^{(\theta)(\phi)}=\dfrac{Q^{(r)(\theta)}w^{(\phi)}}{w^{(r)}},\label{eq:Q_Kerr}  \\
 Q^{(t)(\phi)}&= \dfrac{Q^{(r)(\phi)}w^{(t)}}{w^{(r)}}+\dfrac{w^{(\phi)}}{2 w^{(t)}[w^{(r)}]^2}\bigl\{Q^{(t)(t)}[w^{(r)}]^2-Q^{(r)(r)}[w^{(t)}]^2 \nonumber  \\
 &+Q^{(\theta)(\theta)}([w^{(t)}]^2+[w^{(r)}]^2)\bigr\}. \nonumber
\end{align}
If we now substitute the above relations into $F^{(\mu)}$, we find that 
\begin{align}
    F^{(r)} &=\dfrac{9M\sqrt{\Delta}}{r^5[w^{(r)}]^2}\bigl([w^{(t)}]^2-[w^{(r)}]^2\bigr)\Bigl[Q^{(\theta)(\theta)}([w^{(r)}]^2+[w^{(\phi)}]^2)\nonumber\\
    &+w^{(\phi)}(2Q^{(r)(\phi)}w^{(r)}-Q^{(r)(r)}w^{(\phi)}) -Q^{(\phi)(\phi)}[w^{(r)}]^2\Bigr].\nonumber
\end{align}
and $F^{(t)}=F^{(\theta)}=F^{(\phi)}=0$. The latter result implies because of \ref{eq:MPD_PRR} that $\dot{m}=0$ and, hence, from $F{(r)}=0$, we obtain the following relation
\begin{equation}
Q^{(\phi)(\phi)}=\dfrac{1}{[w^{(r)}]^2}\Big[Q^{(\theta)(\theta)}([w^{(r)}]^2+[w^{(\phi)}]^2)+w^{(\phi)}(2Q^{(r)(\phi)}w^{(r)}-Q^{(r)(r)}w^{(\phi)})\Big].    \label{eq:Qphiphi_Kerr}
\end{equation}
The interpretation we give to the constraints \ref{eq:Q_Kerr} is the same as in Section~\ref{sec:SBH_PDQ}. Namely, in order the pole-dipole-quadrupole body to follow a geodesic trajectory the parallel transported along the geodesic reference vector $w^{(\mu)}$ has to dictate the evolution of the mass quadrupole through these constraints.

\section{Conclusions}

Using trajectories of pole-dipole body that coincide with geodesics on a black hole background, we explored under which condition a pole-dipole-quadrupole body obeying OKS SSC can follow these trajectories as well, when the mass quadrupole tensor is the only component of the quadrupole tensor. For achieving these conditions we have used the not so realistic assumption that the dipole and quadrupole contributions to the MPD vanish independently. We showed that there are constraints between the components of the mass quadrupole tensor and the reference vector $w^\mu$ which have to be obeyed in order the body to follow the aforementioned trajectories. Since the reference vector is parallel transported along a geodesic, these constraints in a sense substitute the missing evolution equations for the quadrupole moment from the MPD equations.

\ack
G.L.-G. and S.M. have been supported by the fellowship Lumina Quaeruntur No. LQ100032102 of the Czech Academy of Sciences.

\bibliography{References}

\end{document}